Generative Diffusion Model-based Downscaling of Observed Sea Surface Height over Kuroshio Extension since 2000

Qiuchang Han[1, 2†], Xingliang Jiang[1, 2†], Yang Zhao[3], Xudong Wang[1, 2*], Zhijin Li[1, 2], and Renhe Zhang[1, 2]

1. Department of Atmospheric and Oceanic Sciences and Institute of Atmospheric Sciences/CMA-FDU Joint Laboratory of Marine Meteorology, Fudan University, Shanghai, China.

2. Key Laboratory of Polar Atmosphere-ocean-ice System for Weather and Climate, Ministry of Education, Fudan University, Shanghai, China.

3. Shenyang Kangtao Technology Co., Ltd., Shenyang, Liaoning, China.

†These authors contribute equally

*Corresponding Author: xdwang_aos@fudan.edu.cn

Abstract

Satellite altimetry has been widely utilized to monitor global sea surface dynamics, enabling investigation of upper ocean variability from basin-scale to localized eddy ranges. However, the sparse spatial resolution of observational altimetry limits our understanding of oceanic submesoscale variability, prevalent at horizontal scales below 0.25° resolution. Here, we introduce a state-of-the-art generative diffusion model to train high-resolution sea surface height (SSH) reanalysis data and demonstrate its advantage in observational SSH downscaling over the eddy-rich Kuroshio Extension region. The diffusion-based model effectively downscales raw satellite-interpolated data from 0.25° resolution to 1/16°, corresponding to approximately 12-km wavelength. This model outperforms other high-resolution reanalysis datasets and neural network-based methods. Also, it successfully reproduces the spatial patterns and power spectra of satellite along-track observations. Our diffusion-based results indicate that eddy kinetic energy at horizontal scales less than 250 km has intensified significantly since 2004 in the Kuroshio Extension region. These findings underscore the great potential of deep learning in reconstructing satellite altimetry and enhancing our understanding of ocean dynamics at eddy scales.

**Keywords:** Diffusion model, Deep learning, Downscaling, Sea surface height, Eddy kinetic energy

# 1. Introduction

Satellite altimeters have significantly improved our understanding of sea surface height (SSH) variations on a global scale over the past four decades. Accurate investigation of upper ocean dynamics relies heavily on satellite altimetry datasets (Fu et al., 2010; MacWilliams 2016; Qiu et al., 2017). A critical limitation of nadir-looking altimeters is the presence of large gaps between satellite tracks (Chelton et al., 2011; Rong and Liang 2022). Generally, along-track SSH can resolve ocean waves with typical wavelengths down to 70 km, while the spatial resolution in two-dimensional gridded SSH map is only on the order of 150 km (Qiu et al., 2017, Ballarotta et al., 2019), inadequate for diagnosing mesoscale and submesoscale processes with horizontal scales less than a few hundred kilometers. Although optimal interpolation, data assimilation, and multi-satellite altimeter data have been utilized to fill the gaps between satellite tracks, the linear interpolation algorithms employed to date still poorly capture the nonlinear relationships between irregular along-track observations and gridded two-dimensional SSH maps (Traon et al., 1998; Ducet et al., 2000; Morrow and Traon 2012; Ubelmann et al., 2015; Fujii et al., 2019; Taburet et al., 2019).

Global and regional ocean models have been developed to improve the horizontal resolution of SSH maps through downscaling, also referred to as super-resolution in the computer vision field. Downscaling through dynamic model simulations is physics-based but comes with substantial computational cost (Kendon et al., 2010; Gudmundsson et al., 2012). Conversely, traditional statistical downscaling methods reduce computational demands but struggle with unlearned data relationships and distributions (Fowler et al., 2007; Sunyer et al., 2012; Maraun et al., 2017). Deep learning-based methods have emerged as a new paradigm for satellite altimetry data downscaling. This approach is data-driven and often outperforms traditional dynamical and statistical downscaling algorithms (George et al., 2021; Manucharyan et al., 2021; Rong and Liang 2022; Fablet et al., 2024; Febvre et al., 2024). Recently, generative diffusion models (DMs) have become a cutting-edge technique for image super-resolution (Ho et al., 2020; Song et al., 2021; Karras et al., 2022; Saharia et al., 2023) and climate downscaling (Addison et al. 2022; Mardani et al. 2023; Bischoff and Deck 2024; Ling et al. 2024; Watt and Mansfield 2024).

The importance of SSH downscaling stems not only from the sparse horizontal resolution of satellite altimetry datasets but also from the profound impacts of ocean eddies on large-scale ocean circulation, heat uptake, marine ecosystem, and climate change. Mesoscale ocean eddies with sizes of 100-250 km contain 80% of the total ocean kinetic energy (Ferrari and Wunsch 2009). In addition, the role of submesoscale processes at scales of 10-100 km has been overlooked due to the low effective resolution of satellite observations. Using high-resolution ocean simulations, recent studies have suggested that submesoscale processes are crucial for upper ocean heat transport (McWillaims 2016; Su et al. 2018; Zhang et al. 2023) and El Nino-Southern Oscillation (Wang et al. 2022).

In this study, we developed a state-of-the-art generative DM to successfully downscale traditional AVISO observational SSH data from 2000 onwards in the eddy-rich Kuroshio Extension region, which is an ideal test-bed for high-resolution SSH mapping. Despite the lack of observational high-resolution SSH data, we leverage model reanalysis outputs to extract fine-scale SSH information from coarse satellite observations. Significantly, our newly generated dataset reveals an intensification trend in eddies over the recent two decades in the Kuroshio Extension region.

## 2. Results

### 2.1 Diffusion model-based downscaling using ocean reanalysis

Reanalysis products employ state-of-the-art ocean models incorporating data assimilation schemes to produce SSH that closes to observations (Carrassi et al., 2018; Lellouche et al., 2020). We first compare the GLORYS12 reanalysis to AVISO observations (see Methods). AVISO captures large-scale ocean circulation; however, due to its coarse resolution, ocean eddies with relative vorticity comparable to Earth's rotation rate ($|Ro| > 0.1$; see Methods) typically appear on the map as isolated centers (Figs. 1a and 1e). In contrast, GLORYS12 well reproduces the observed large-scale patterns while simulating wavy characteristics and filaments of ocean eddies (Figs. 1b and 1f). Cyclonic and anticyclonic eddies are ubiquitous over the Kuroshio Extension region in daily snapshots. On January 15, 2022, AVISO reveals an array of cyclonic and anticyclonic eddies with a southwest-northeast orientation east of landmass (Fig. 1c). While GLORYS12 indeed reproduces these observed eddies, it also spuriously generates northwest-southeast oriented filaments that are absent in the observations for that day (Fig. 1d). Similar discrepancies are evident on April 1, 2022, where the cyclonic submesoscale eddy simulated by GLORYS12 is displaced southward (Fig. 1h), inconsistent with AVISO observations (Fig. 1g).

Given the aforementioned discrepancies, GLORYS12 can not be treated as the observational ground truth. However, the eddy-scale variability it generates is dynamically constrained by model simulations. Also, its large-scale variability closely resembles observations. Thus, we propose a SSH downscaling scheme for the Kuroshio Extension region based on generative DMs using GLORYS12 reanalysis data, which we term the Kuroshio Extension SSH Downscaling Diffusion Model (KESHDiff, see Methods).

DMs contains a forward diffusion process and a reverse denoising process (Fig. 2a) (Ho et al., 2020; Song et al., 2021). As generative models, DMs aim to learn probability distribution $p(\mathbf{x})$ given a set of samples $\{\mathbf{x}\}$, with standard deviation $\sigma_{data}$. The forward diffusion process gradually adds varying levels of independent identically distributed Gaussian noise of standard deviation σ to the data to get $p(\mathbf{x}; \sigma)$. For $\sigma_{max} \gg \sigma_{data}$, $p(\mathbf{x}; \sigma_{max})$ is practically similar with pure Gaussian noise, as illustrated by the images from $\mathbf{x}(0)$ to $\mathbf{x}(T)$ in Fig. 2a. For the reverse process, the DMs sequentially denoises $\mathbf{x}(T)$ back into $\mathbf{x}(0)$, such that at each noise level $\mathbf{x}(i) \sim p(\mathbf{x}_i; \sigma_i)$. The endpoint of the reverse process is thus distributed according to the learned data distribution. The above forward diffusion and reverse denoising processes are unconditional, meaning that the generated results are entirely based on the probability distribution learned from the training data. . In our work, we focus on conditional downscaling with our proposed model, KESHDiff. While KESHDiff's forward diffusion process is similar to other well-developed DMs, its reverse denoising process incorporates low-resolution data as additional conditions to guide the generation of high-resolution counterparts (Fig. 2b). This conditional approach allows for more controlled and targeted generation of high-resolution data.

To solve the reverse denoising process in KESHDiff, we utilize a U-Net-like architecture (Fig. 2b, for details see Methods).

## 2.2 Model comparison on the GLORYS12 validation set

We comprehensively compare KESHDiff against three other models, including bilinear interpolation, super-resolution based on Generative Adversarial Network (SR-GAN), and U-Net, using the validation set of GLORYS12 (see Methods). KESHDiff outperforms other models across six metrics (Fig. 3 and Supplementary Table 1), including Peak Signal-to-Noise Ratio (PSNR), Structure Similarity Index Measure (SSIM), Mean Absolute Error (MAE), Root Mean Square Error (RMSE), Temporal Correlation Coefficient (TCC), and Pattern Correlation Coefficient (PCC). Specifically, KESHDiff achieves a PSNR exceeding 53dB, while bilinear interpolation also surpasses 50dB, indicating that traditional statistical downscaling methods still retain some efficacy (Supplementary Table 1). However, well-trained neural networks like U-Net and KESHDiff demonstrate superior performance across multiple metrics. U-Net outperforms bilinear interpolation in terms of SSIM and PCC. In contrast, SR-GAN yields inferior results (Fig. 3).

Note that for 2-dimensional fields, PSNR, MAE, and RMSE are metrics that quantify the average error across all wavelengths (or scales), thereby focusing on the overall error rather than the error at specific wavelengths. In image signal processing, the long-wavelength components dominate the overall quality of the image and the aforementioned metrics, with the influence of short-wavelength components being almost negligible. However, SSH signals are different, as signals at various wavelengths carry substantial dynamical importance, requiring further assessment in the Fourier space (see Methods). Figure 4a shows the power spectral density (PSD) of high-resolution SSH from GLORYS12, serving as the ground truth, alongside the SSH outputs from four methods. The power spectrum of KESHDiff is indistinguishable from the ground truth across all wavelength, indicating that the model has effectively learned the spectral characteristics of SSH. The shortest wavelength in the high-resolution data is approximately 12 km, as shown on the x-axis of Fig. 4a. Given that we are implementing an eight-fold downscaling scheme (see Methods), the shortest identifiable wavelength for the input low-resolution data (Fig. 2b) is only about 96 km. Consequently, the SSH data with wavelengths ranging from 96 km to 12 km are all generated by the models themselves. At wavelengths larger than 500 km, U-Net overestimates the large-scale longwave features, while bilinear interpolation underestimates the energy of longwaves. In the wavelength range from 500 km to 60 km, bilinear interpolation, SR-GAN, and U-Net underestimate the energy. At wavelengths shorter than 60 km, bilinear interpolation performs consistently, whereas the other two neural network-based downscaling schemes inappropriately generate more perturbations.

We also examine the error ratios between these models and the ground truth. The error ratios represent the ratio between the spectra of error and spectrum of the ground truth across different frequency bands (see Methods). From large-scale down to approximately 96 km, the error ratios of three models, excluding SR-GAN, are all less than 0.5, suggesting good performance in capturing large-scale SSH characteristics with relatively small errors (Fig. 4b). Among these, KESHDiff exhibits the lowest error ratio, closely resembling the ground truth. In the model-generated frequency bands, specifically the 96 km to 12 km range, KESHDiff consistently maintains an error ratio below 0.5. U-Net performs well in the 96-20 km range, but its error rapidly increases in the shorter wavelength band, indicating artificially enhanced power spectrum energy in this band (Fig. 4a). Bilinear interpolation exhibits an even lower ratio than KESHDiff in the 20-12km range, likely due to smoothing data between grid points, resulting in low power spectrum energy (Fig. 4a), rather than being driven by reasonable physical processes.

**2.3 Downscaling AVISO SSH utilize KESHDiff**

KESHDiff exhibits exceptional performance in SSH downscaling on the GLORYS12 validation set. Thus, we utilize KESHDiff to downscale the gridded AVISO SSH from a horizontal resolution of $0.25°$ to $1/16°$, achieving a shortest wavelength of approximately 12 km (see Methods). We compare our KESHDiff output with two sets of high-resolution ocean model reanalysis data assimilating observations (HYCOM and GLORYS12), the original $0.25°$ low-resolution AVISO data used as input, and multi-satellite along-track observations (see Methods for data descriptions). The comparison period spans from December 1, 2020, to December 31, 2022.

The PSD of the along-track observations gradually decreases with shorter wavelengths (Fig. 5a), a feature shown in all datasets. HYCOM data performs the worst, exhibiting significant errors compared to the along-track observations at wavelengths exceeding 300 km (Fig. 5b). Among the remaining datasets, the discrepancy between KESHDiff and the along-track observations is smallest (Fig. 5b). Surprisingly, although our training set is derived from GLORYS12, our model outperforms GLORYS12 itself when applied to AVISO data. In addition, KESHDiff is also better than the input low-resolution AVISO SSH on large scales. This indicates that ocean reanalysis serves as an effective proxy for direct high-resolution SSH observations in the model training step, despite unavailability of the latter.

At scales below 300 km, due to the coarse effective resolution, AVISO lacks structures at the corresponding scales, leading to a rapid decline in its PSD (Fig. 5a). Besides, the two model reanalyses show consistent declines in PSD compared to KESHDiff. In contrast, KESHDiff remains the closest to the along-track observations even at the shortest effective wavelength of 70 km (Fig. 5b).

Consistent with Fig. 1, daily snapshots further demonstrate that KESHDiff has successfully downscaled the SSH over the Kuroshio Extension without generating spurious eddy-scale features (Supplementary Fig. 1).

**2.4 Intensification of eddy variability since 2004**

Based on the output high-resolution SSH data generated by KESHDiff, we focus on the long-term changes in eddy-scale EKE over the Kuroshio Extension region. Area averaged EKE at wavelengths shorter than 250 km (see Methods) exhibits distinct interannual variations (Fig. 6). The EKE was relatively weak in 2015-2016, following the super El Nino, while it intensified during the triple-dip La Nina event of 2020-2022. The 23-year average EKE from 2000 to 2022 is 490.82 $cm^2/s^2$ (Fig. 6). Since 2000, the EKE at eddy scales shows a slight but insignificant decreasing trend (see Methods). However, EKE has significantly intensified since 2004 at a rate of 10.14 $cm^2/s^2$ per decade, indicating an increase of 2.07% per decade. The intensification of EKE is even more rapid since 2007, reaching 23.74 $cm^2/s^2$ per decade (4.84% per decade).

**3. Discussion**

Despite significant improvement of satellite observations since the 1980s, the sparse satellite altimetry data still limits our understanding of ocean surface eddies. The new Surface Water and Ocean Topography (SWOT) satellite, launched in late 2022, can directly observe ocean submesoscale processes (Durand et al., 2010). However, its coverage period is still too short to investigate historical ocean variability. On the other hand, as an promising method, deep learning has already seen widespread application in the climate and Earth sciences. Studies utilizing generative diffusion models for SSH downscaling have only emerged recently (Li et al., 2024). Based on a state-of-the-art generative diffusion model, we have developed an SSH downscaling algorithm for the Kuroshio Extension region, an ideal test-bed for high-resolution SSH mapping. We refer to this algorithm as KESHDiff.

KESHDiff displays incredible superiority in SSH downscaling over the Kuroshio Extension region. Firstly, it outperforms current widely-used statistical downscaling and neural network-based downscaling algorithms across six evaluation metrics (Fig. 3). Secondly, it exhibits exceptional performance on the validation set of high-resolution ocean model reanalysis data, with spectral features closely matching model outputs. In addition, when applied to observational gridded low-resolution AVISO SSH data, it proves to be the closest to the along-track observations, even surpassing both the input AVISO data and the GLORYS12 reanalysis data used for model training.

Last, using the high-resolution SSH data reconstructed by KESHDiff for 2000-2022, we revealed a linear intensification trend in EKE over the Kuroshio Extension region at horizontal scales less than 250 km. Since 2004, ocean surface eddy-scale variability has intensified by 10.14 $cm^2/s^2$ (2.07%) per decade, consistent with previous conclusions (Martinez-Moreno et al., 2021).

This study shows great implication for future deep learning-associated satellite oceanography research. While we focus on the Kuroshio Extension region, data-driven deep learning downscaling algorithms are applicable to any ocean areas, differing with traditional downscaling methods (Martin et al., 2024). Furthermore, deep learning algorithms require less computational cost compared to traditional ocean numerical simulations. For instance, the training stage of KESHDiff utilized one NVIDIA RTX 3090 GPU. After training, generating daily high-resolution SSH data for one year takes only 3-4 hours. These conclusions do not imply that deep learning-based methods have surpassed traditional ocean models. KESHDiff still heavily relies on high-quality high-resolution ocean reanalysis data, highlighting that current deep learning-based ocean downscaling and prediction methods require the support from dynamical models (Zheng et al., 2020).

Finally, more detailed conclusions of long-term changes of ocean surface eddy variability need further investigation.

## Methods

### Datasets

For gridded satellite observations, we use daily SSH from the Archiving, Validation and Interpretation of Satellite Oceanographic data (AVISO) of the Copernicus Marine Environment Monitoring Service (CMEMS) with a horizontal resolution of 0.25° from 2000 to 2022.

For satellite along-track SSH observations, we use multiple satellites passing over the Kuroshio Extension region. The data is partly processed by the Data Unification and Altimeter Combination System (DUACS) multimission altimeter data processing system. The satellites include Jason 3, Sentinel 3A, Sentinel 3B, HY 2B, and Sentinel 6 in DUACS. Specifically, we use the unfiltered, Level 3 sea level anomaly observations. At Level 3, the observations have been corrected for atmospheric effects, the barotropic tide has been removed, and the data has been adjusted to ensure consistency between the different altimeter missions. Satellite along-track observations cover the period from December 1, 2020 to December 31, 2022.

For high-resolution SSH reanalysis datasets, we used the daily SSH from the CMEMS global ocean eddy-resolving (1/12º horizontal resolution) reanalysis (GLORYS12) covering the period from July 1, 2020 to January 31, 2024 (Jean-Michel et al., 2021). We further utilized the data-assimilative HYbrid Coordinate Ocean Model (HYCOM) with a horizontal resolution of 1/12º (Chassignet et al., 2007).

In this study, the Kuroshio Extension region is defined as the area between 27ºN-41.22ºN, 140ºE-168.44ºE. To facilitate SSH downscaling, we interpolate the GLORYS12 and HYCOM high-resolution reanalyses to a 1/16º resolution using bicubic interpolation.

**The kernels of diffusion models**

As mentioned in Section 2.1, the generative DMs were inspired from non-equilibrium thermodynamics (Sohl-Dickstein et al., 2015). The model contains forward diffusion process and reverse denoising process. The forward diffusion process transforms samples from the data distribution $\mathbf{x}(0) \sim p(\mathbf{x}; \sigma_{data})$ to that of pure Gaussian noise $\mathbf{x}(T) \sim N(0, \sigma I)$, which can be represented by the stochastic differential equation (SDE). Following Song et al. (2021), the SDE for such a process is given by

$$d\mathbf{x} = f(\mathbf{x}, t)dt + g(t)dw$$

Where $f(\mathbf{x}, t)$ is the drift coefficient, $g(t)$ is the diffusion coefficient, and w is a Wiener process. If the SDE can be reversed, we can use it to generate samples from $p(\mathbf{x}(0))$. The reverse denoising SDE process is:

$$d\mathbf{x} = [f(\mathbf{x}, t) - g(t)^2 \nabla_{\mathbf{x}} log p_t(\mathbf{x})]dt + g(t)d\overline{w}$$

Where $\overline{w}$ is a Wiener process with time flowing backwards from T to 0. $\nabla_{\mathbf{x}} log p_t(\mathbf{x})$ is the score function, a vector denotes the gradient of the log probability density function of the data distribution.

We can also begin by examining the deterministic ordinary differential equation (ODE). The reverse ODE process is then as follows:

$$d\mathbf{x} = [f(\mathbf{x}, t) - \frac{1}{2} g(t)^2 \nabla_{\mathbf{x}} log p_t(\mathbf{x})]dt$$

Following Karras et al. (2022), the ODE can also be written as:

$$dx = [\frac{\dot{s}(t)}{s(t)}\mathbf{x} - s(t)^2\dot{\sigma}(t)\sigma(t)\nabla_\mathbf{x} log p(\frac{\mathbf{x}}{s(t)};\sigma(t))]dt$$

Where $s(t)$ is scaling factor. Thus, the reverse ODE mainly focuses on the reparameterization of $s(t)$ and $\sigma(t)$. In addition, unlike traditional DMs (i.e. DDPM and DDIM), the diffusion and denoising processes can be simplified as the adding or subtracting Gaussian noise with varying intensities (different standard deviations; Karras et al., 2022). Thus, $\sigma(t) = t$ in our KESHDiff model. We use 182 time steps during evaluation, but find that using just 100 is sufficient for a similar accuracy.

**The kernels of U-Net**

The structure of the U-Net is shown in Figure 2b. To improve efficiency of the U-Net, rather than directly learning the output on the fine resolution grid, we learn the difference between the output on the fine and coarse grid. We expect these to aid learning of details around the coast. This means the input image is of size ($256\times512\times2$), while the output image is of size ($256\times512\times1$). The U-Net is trained by minimizing the Mean Squared Error (MSE) between the U-Net predicted image and the samples from the data.

**Super-Resolution Generative Adversarial Network (SR-GAN)**

The SR-GAN network is consistent with previous paper (Ledig et al., 2017). The VGG19 network used in perceptual loss is mainly for extracting high-level image features to better measure the similarity between the generated image and the original high-resolution image.

**Kuroshio Extension SSH downscaling diffusion model**

Since we utilize a conditional diffusion model, an additional input of a low-resolution image is required for SSH downscaling (Fig. 2b). However, after interpolating the 1/16º GLORYS12 data to 1/4º, the resulting images still contain more details than the original 1/4º AVISO data. Thus, the 1/16º high-resolution GLORYS12 data is first spline-interpolated to a 0.5º low-resolution data. These high and low-resolution GLORYS12 datasets are then included in the training set to develop an eight-fold downscaling model. For the Kuroshio Extension region, a 1/16º resolution corresponds to 512 grid points in the zonal direction and 256 grid points in the meridional direction. The 0.5º resolution thus should be 64 grid points in the zonal direction and 32 grid points in the meridional direction.

The sample time range for the training set is from July 1, 2020, to July 31, 2023, while the validation set includes March 2018, June 2019, September 2023, and January 2024. The sample time range for the validation set is not included in the training set.

When applying KESHDiff to AVISO, we first spline-interpolate the AVISO data to a 0.5° resolution, then perform eight-fold downscaling to 1/16°. The AVISO outputs cover the period from January 1, 2000, to December 31, 2022.

**Metrics for model comparison**

We calculate six metrics to compare and evaluate different downscaling algorithms, including Peak Signal-to-Noise Ratio (PSNR), Structure Similarity Index Measure (SSIM), Mean Absolute Error (MAE), Root Mean Square Error (RMSE), Temporal Correlation Coefficient (TCC), and Pattern Correlation Coefficient (PCC). Their definitions are as follows:

$$PSNR = \frac{1}{n}\sum_{i=1}^{n} 10 \times log_{10}\left(\frac{C_{max}^2}{MSE}\right)$$

$$MSE = \frac{1}{n}\sum_{i=1}^{n}\left(h_i - h_i'\right)^2$$

$$SSIM = \frac{1}{n}\sum_{i=1}^{n}\frac{\left(2\mu_i\mu_i' + C_1\right)\left(2\sigma_i\sigma_i' + C_2\right)}{\left((\mu_i)^2 + (\mu_i')^2 + C_1\right)\left((\sigma_i)^2 + (\sigma_i')^2 + C_2\right)}$$

$$MAE = \frac{1}{n}\sum_{i=1}^{n}\left|h_i - h_i'\right|$$

$$RMSE = \sqrt{\frac{1}{n}\sum_{i=1}^{n}\left(h_i - h_i'\right)^2}$$

Here, n represents the total number of sample days in the validation set. $h_i$ and $h_i'$ denote the GLORYS12 high-resolution reanalysis and the downscaling outputs, respectively. $\mu_i$ and $\mu_i'$ represent the spatial means of $h_i$ and $h_i'$, while $\sigma_i$ and $\sigma_i'$ represent spatial standard deviations. $C_{max}$ denotes the maximum value of the daily SSH, and equals to 3. $C_1$ and $C_2$ are constants to avoid computation instability when the denominator approaches zero.

TCC is defined as the temporal correlation coefficient of $h_i$ and $h_i'$ at the same non-land grid points, followed by averaging the correlation coefficients across these non-land points. PCC is the spatial correlation coefficient of $h_i$ and $h_i'$ for each day, followed by averaging over n days.

To plot Fig. 3, we further normalize the modified metrics:

$$PSNR_{mod} = 0.5 + 0.5 * \frac{PSNR - PSNR_{min}}{PSNR_{max} - PSNR_{min}}$$

$$SSIM_{mod} = \frac{SSIM - 0.95}{SSIM_{max} - 0.95}$$

$$1 - MAE_{mod} = 0.5 + 0.5 * (1 - \frac{MAE - MAE_{min}}{MAE_{max} - MAE_{min}})$$

$$1 - RMSE_{mod} = 0.5 + 0.5 * (1 - \frac{RMSE - RMSE_{min}}{RMSE_{max} - RMSE_{min}})$$

$$TCC_{mod} = \frac{TCC - 0.99}{TCC_{max} - 0.99}$$

$$PCC_{mod} = \frac{PCC - 0.99}{PCC_{max} - 0.99}$$

Where max and min represent the maximum and minimum values for each metric across four models.

**Two-dimensional (2D) Fourier transforms and radial averaging of power spectrum**

To calculate the power spectra of various datasets, we utilize a 2D discrete fast Fourier transform (2DFFT). To focus on the regional scale variability, we first kick off the land points, followed by subtracting the zonal mean from the SSH data. As suggested by Sasaki et al. (2014), the zonal mean values affect scales larger than 300 km.

The calculation steps for Fig. 4 are as follows: We apply 2DFFT to the daily SSH for the sample period to obtain the power spectra, and then average the daily power spectra over the entire sample period. The 2DFFT converts SSH from the spatial domain into the Fourier frequency domain. The 2DFFT of SSH $h(x, y)$ is defined as:

$$H(X, Y) = \sum_{x=0}^{M-1} \sum_{y=0}^{N-1} h(x, y) \cdot e^{-i2\pi(\frac{ux}{M} + \frac{vy}{N})}$$

Where:

$h(x, y)$ is the input SSH data in the spatial domain. $H(X, Y)$ is the output signal in the frequency domain. M and N are the dimensions of the input SSH data. $X$ and $Y$ are the frequency components corresponding to the spatial dimensions x and y. $i$ is the imaginary unit.

The power spectral density (PSD) represents the distribution of power across different frequency components. The PSD $P(X, Y)$ is given by:

$P(X, Y) = |H(X, Y)|^2$

Where $|H(X, Y)|$ is the magnitude of the Fourier coefficients.

To reduce the 2D PSD to a one-dimensional function, radial averaging is performed. This process involves averaging the PSD over concentric circles of constant radius $r$ in the frequency domain. The radius $r$ is defined as:

$$r = \sqrt{X^2 + Y^2}$$

The radially averaged PSD is thus calculated by averaging the $P(X, Y)$ over all points $(X, Y)$ that lie on the circle with radius $r$.

After doing the above steps, the power spectra for multi days are averaged to produce the final power spectrum used for analysis and plotting in Figs. 4 and 5.

**Definition of error ratios in spectral analysis**

In this study, we subtract the daily high-resolution SSH outputs of each model on the validation set from the high-resolution GLORYS12 data, then apply 2DFFT to obtain the power spectrum of the error, and finally divide it by the power spectrum of the GLORYS12 high-resolution data. Therefore, the error ratio indicates the differences between the downscaling model and the input ground truth across different frequencies. Empirically, a ratio greater than 0.5 represents that the downscaling model has significant errors, whereas a ratio less than 0.5 suggests that the model well resembles the ground truth.

**Comparison between high-resolution SSH datasets with along-track observations**

For the comparison of power spectra between various datasets and the satellite along-track observations, we only consider data in the direction of the satellite nadir point.

**Definition of eddy-scale variability**

In this study, we define eddy scale variability, encompassing both mesoscale and submesoscale, by high-pass filtering with a cut-off wavelength of 250 km via 2DFFT.

**Calculation of the Rossby number $Ro$**

Relative vorticity $\xi$ is defined as the rotational component of horizontal motions:

$$\xi = \frac{\partial v}{\partial x} - \frac{\partial u}{\partial y}$$

Where $u$ and $v$ are the zonal and meridional velocities.

To derive the $\xi$ from SSH, the $u$ and $v$ are obtained assuming geostrophic balance under the approximation:

$$u_g = -\frac{g}{f}\frac{\partial h}{\partial y}$$

$$v_g = \frac{g}{f}\frac{\partial h}{\partial x}$$

Where $g$ is the acceleration due to gravity, and $f$ is the Coriolis parameter (Earth's rotation rate). The relative vorticity should be:

$$\xi = \frac{g}{f}(\frac{\partial^2 h}{\partial x^2} + \frac{\partial^2 h}{\partial y^2})$$

The Rossby number Ro is a non-dimensional number is utilized to represent the surface frontal structure (Su et al., 2018). It is defined as: $Ro = \xi/f$. When this non-dimensional number is larger than 0.1, the ocean eddies may be large (Sasaki et al., 2014; Klein et al., 2008; Mensa et al., 2013).

**Eddy kinetic energy**

The kinetic energy, KE, is calculated from the surface geostrophic current maps

$$KE = \frac{1}{2}(u_g^2 + v_g^2)$$

The eddy kinetic energy, EKE, is defined as the time-varying component of the KE using a Reynolds decomposition:

$$EKE = \frac{1}{2}(\overline{u'^2} + \overline{v'^2})$$

Where the constant ocean water density is ignored. $u'$ and $v'$ represent the eddy-scale velocities via a 2DFFT.

**Trends and significance**

The KESHDiff outputs the high-resolution daily SSH data over 2000-2022. We first calculate the monthly mean EKE and then obtain the time series over 2000-2022 by averaging the monthly EKE over the region 33°N-40°N, 142°E-158.5°E. Then, we apply a 35-month (3-year) running average to remove the effects of seasonality and high-frequency variability.

On the basis of the 3-year running mean time series, we calculate the trends and statistical significance using the Theil-Sen estimator (Mondal et al., 2012) and the Mann-Kendall test (Wang et al., 2024).

**Data Availability**

All satellite and ocean model reanalysis data used in this study are publicly available. The AVISO SSH is available at https://resources.marine.copernicus.eu/product-detail/SEALEVEL_GLO_PHY_L4_NRT_OBSER

VATIONS_008_046/DATA-ACCESS. The GLORYS12V1 is available at https://data.marine.copernicus.eu/product/GLOBAL_MULTIYEAR_PHY_001_030/description. The HYCOM is available at https://www.hycom.org/dataserver. We followed the diffusion implementation of Karras et al., 2022 and Watt and Mansfield 2024, available at https://github.com/NVlabs/edm and https://github.com/robbiewatt1/ClimateDiffuse, respectively. We will also be releasing our code on GitHub and sharing our dataset soon.

## Acknowledgements

Q. H., X. J., X. W., Z. L., and R. Z. at Fudan University were supported by the National Natural Science Foundation of China (42288101). X. W. was also supported by the National Natural Science Foundation of China (42205016). We appreciate Dr. Fenghua Ling's suggestions in the early stages of this paper.

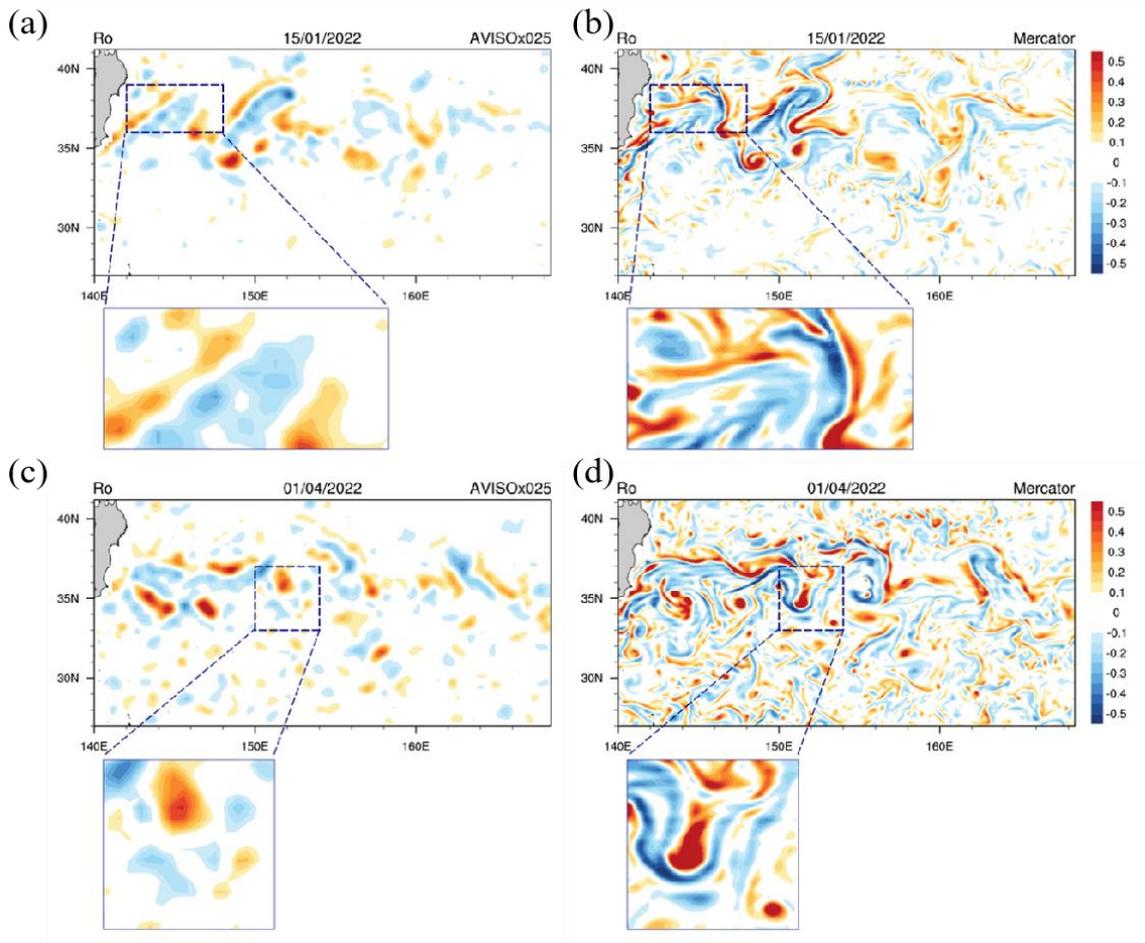

**Fig. 1. Surface Rossby number (Ro=ξ/f; shading) in the KOE region.** (a and b) On 15 January 2022 (c and d) and on 01 April 2022. Ro in the AVISO dataset is shown in (a and b), while (b and d) represent the Ro in GLORYS12 reanalysis.

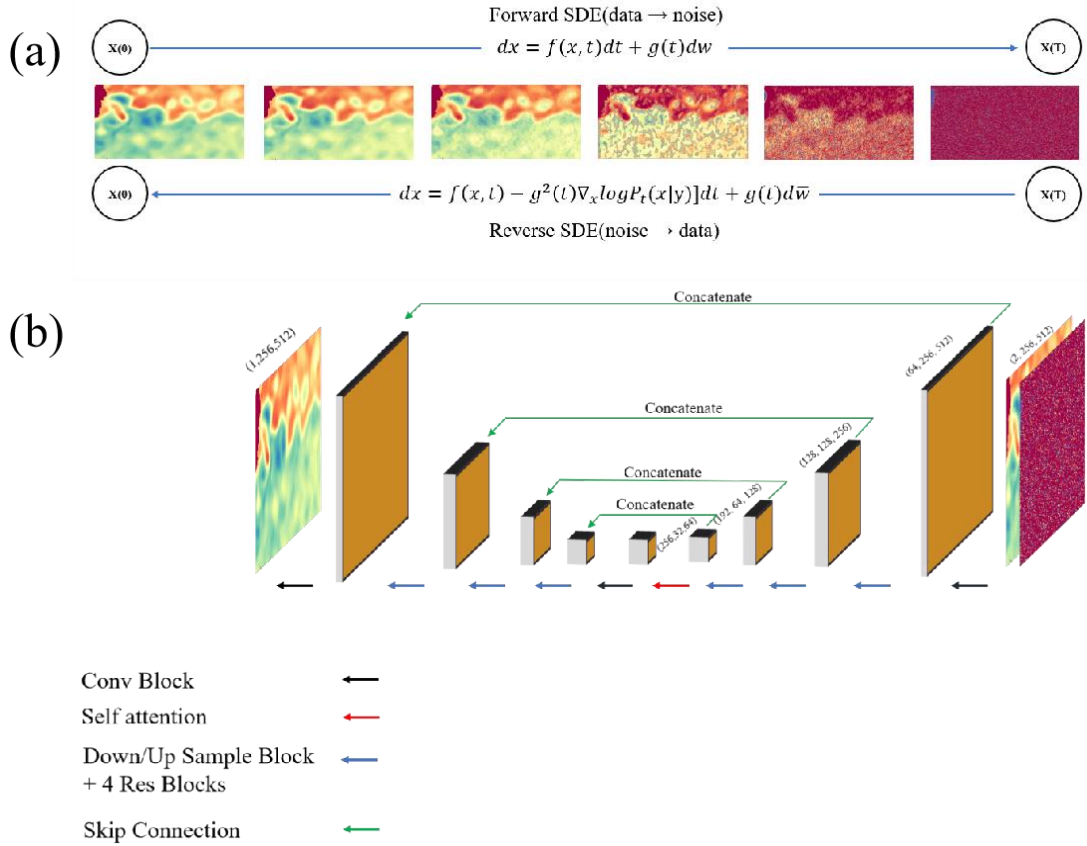

**Fig. 2. The structure of the Kuroshio Extension SSH downscaling diffusion model (KESHDiff).** (a) The forward diffusion process from 0 to T, and the reverse denoising process from T to 0. (b) Architecture of the U-Net applied to KESHDiff. In the reverse denoising process, low-resolution data is provided as input. The U-Net architecture contains both upsampling and downsampling use similar residual modules and self-attention modules.

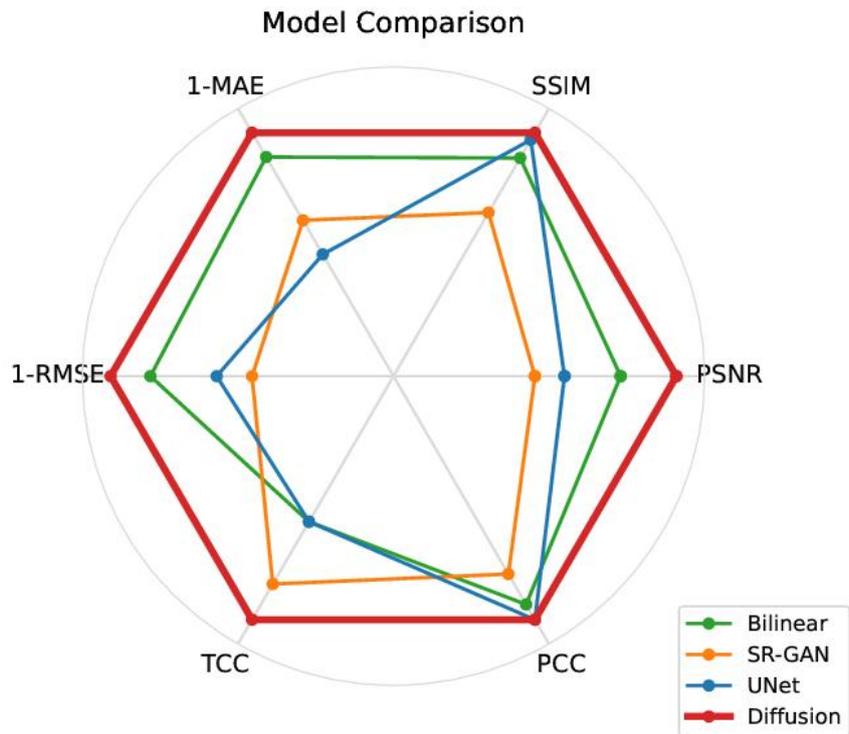

**Fig. 3. Comparison of models across six metrics.** The six modified metrics (see Methods) used in this study including PSNR, SSIM, 1-MAE, 1-RMSE, TCC, and PCC. Legends in the right corner represent four different models. The closer to the outer gray circle, the better the model's performance.

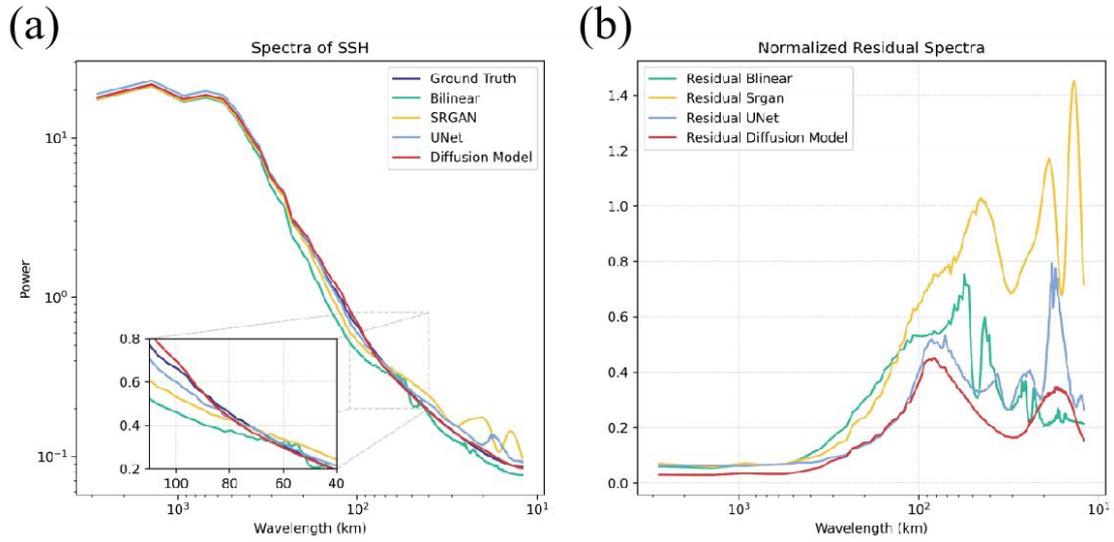

**Fig. 4. Power spectra of different models on the GLORYS12 validation set.** (a) Power spectra of KESHDiff (red), U-Net (light blue), SR-GAN (yellow), bilinear interpolation (green), and the GLORYS12 high-resolution data (blue) via a 2DFFT. (b) The error ratios of spectra between different models and GLORYS12 high-resolution (see Methods).

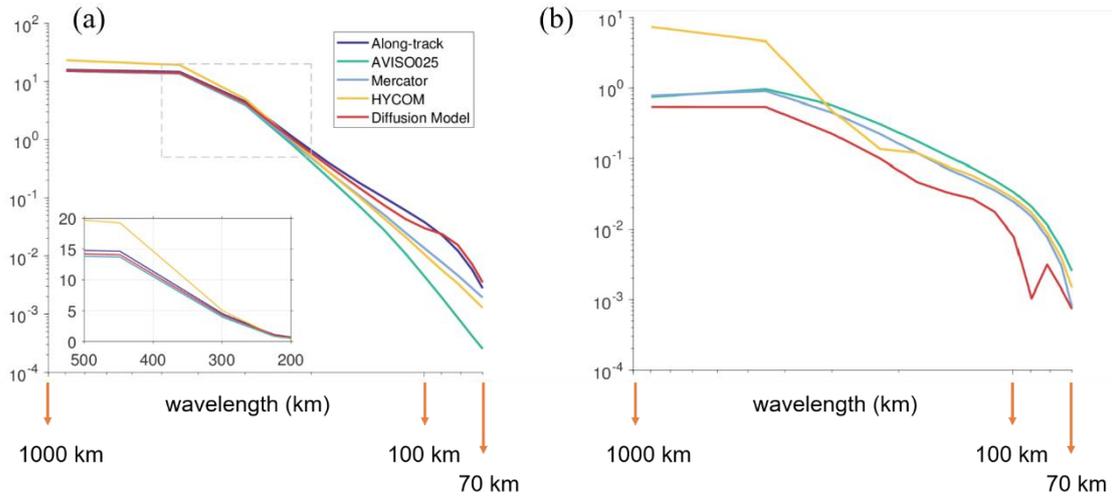

**Fig. 5. Power spectra of different output datasets.** (a) Power spectra of 1/16º resolution KESHDiff (red), 1/16º resolution HYCOM (yellow), 1/16º resolution GLORYS12 (light blue), 1/4º resolution AVISO input (green), and multi-satellite along-track observations (blue) via a 2DFFT. (b) The difference between four gridded datasets and long-track observations in (a).

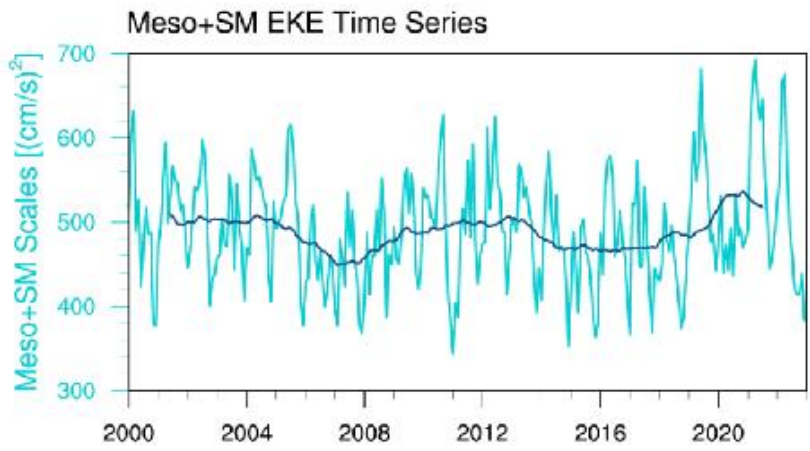

**Fig. 6. Time series of monthly EKE over 2000-2022.** monthly EKE at eddy-scale (light blue) and its 35-month running averaged component (blue curve).

Supplementary Table 1. Performance of four models across six metrics

|        | PSNR(dB) | SSIM   | MAE(m) | RMSE   | TCC    | PCC    |
|--------|----------|--------|--------|--------|--------|--------|
| Bilinear | 50.4494 | 0.9896 | 0.0020 | 0.0030 | 0.9949 | 0.9990 |
| SR-GAN | 45.8022 | 0.9797 | 0.0033 | 0.0053 | 0.9970 | 0.9978 |
| UNet   | 47.4044 | 0.9929 | 0.0040 | 0.0045 | 0.9949 | 0.9996 |
| Diffusion | 53.4701 | 0.9942 | 0.0015 | 0.0021 | 0.9982 | 0.9996 |

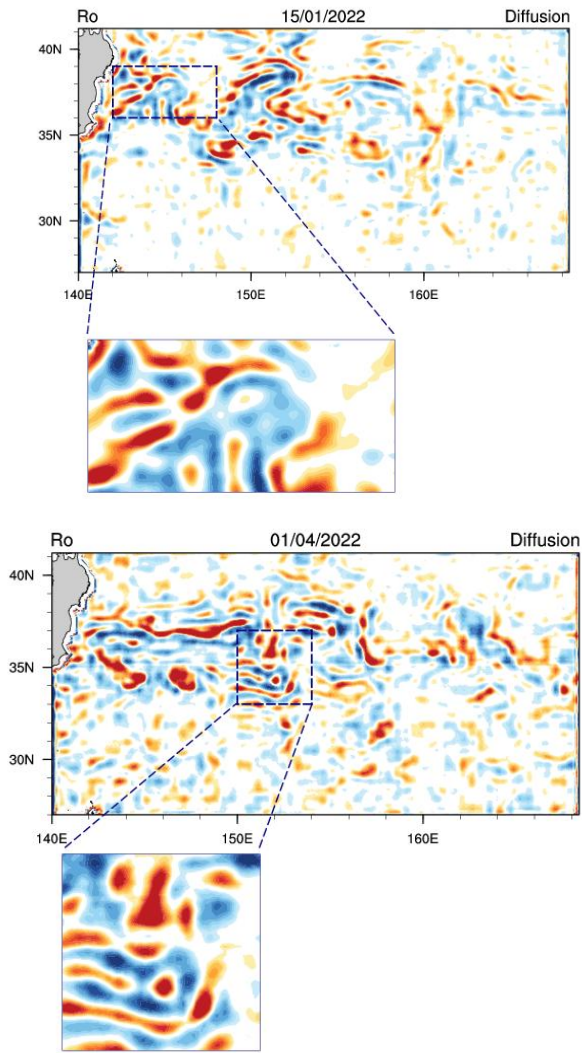

Supplementary Fig. 1. Same as Fig. 1, but for daily snapshots of 1/16º resolution KESHDiff SSH output. Top panel shows Rossby number Ro on January 15, 2022, while bottom panel shows on April 1, 2022.